\documentclass[twocolumn,amsmath,amssymb]{revtex4}

\usepackage{graphicx}
\usepackage{bm}

\begin{document}

\title{Realization of SOC behavior in a dc glow  discharge plasma}%
\author{Md.Nurujjaman}
\email{md.nurujjaman@saha.ac.in}

\author{A.N.Sekar Iyengar}

\affiliation{Plasma Physics Division,
Saha Institute of Nuclear Physics,
1/AF, Bidhannagar,  
Kolkata -700064, India.}
\begin{abstract}
Experimental observations consistent with Self Organized Criticality (SOC) have been obtained in the electrostatic floating potential fluctuations of a dc glow discharge plasma. Power spectrum exhibits a power law which is compatible with the requirement for SOC systems. Also the estimated value of the Hurst exponent (self similarity parameter), H being greater than 0.5, along with an algebraic decay of the autocorrelation function, indicate the presence of temporal long-range correlations, as may be expected from SOC dynamics. This type of observations in our opinion has been reported for the first time in a glow discharge system. 
\end{abstract}
\maketitle
\section{Introduction}

The study of nonequilibrium phenomena  of naturally occurring~\cite{book:kaw}, and laboratory plasmas~\cite{book:kaw,POP:BACarreras0} has been an active area of research in plasma physics. Within this frame work, the  Self Organized Criticality (SOC) concept has been quite rigorously deployed to explain some of the turbulent transport observations in magnetically confined fusion devices like Tokamaks~\cite{POP:BACarreras0,PRL:FSattin}. The physics of glow discharge plasmas~\cite{book:von} in the last two decades, have generated a renewed interest due to their importance in low temperature plasma applications ~\cite{prl:chu,JPhysD:Roth}. Being  a nonlinear medium, they have been a good test bed to investigate various nonlinear phenomena like chaos etc~\cite{POP:Bruhn,PRAM:jaman,PRE:Ding}. In this paper we have attempted  to apply, quite successfully the SOC concept for the first time to some of the turbulent fluctuations in a glow discharge plasma.

The outline of the paper is as follows: In Section~\ref{sec:Setupresult} we describe the experimental setup and observations. In Section~\ref{sec:SOC} we present the analysis techniques and results of the spectral and statistical methods that have been carried to substantiate the SOC behavior. Finally we summarize our results in Section~\ref{sec:conclusion}.  

\section{Experimental Setup and Results}
\label{sec:Setupresult}
\begin{figure}[h]
\center
\includegraphics [width=8.5cm] {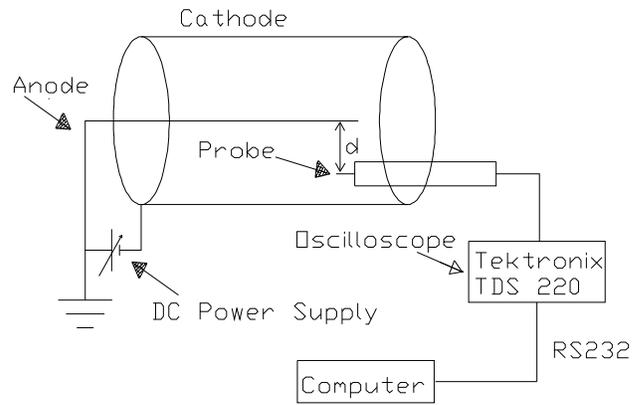}
\caption{Schematic diagram of the experimental setup of the cylindrical dc discharge plasma system with Langmuir probe.}
\label{fig:1}
\end{figure}

The experiment was carried out in a coaxial cylindrical  glow dc discharge plasma system
with argon as shown in Figure ~\ref{fig:1}. The hollow Stainless Steel (SS) outer cylinder
of 45 mm diameter is the cathode and the SS rod of 1 mm diameter inside the 
cathode is the anode, which is grounded. The whole system was placed in
a vacuum chamber, and evacuated to a base pressure of $10^{-3}$ Torr by means of a
rotary pump. Argon gas was introduced using precision needle valve into the chamber.
Plasma discharges were obtained over a wide range of filling pressure and discharge voltage. A Langmuir probe made of tungsten of diameter $\approx 0.5$ mm and length $\approx 2$ mm was used to measure the electrostatic floating potential fluctuations in the plasma at about 12.5 mm from the anode (Fig.~\ref{fig:1}). The fluctuating data of 2500 points was recorded at a sampling rate $\approx 10^{-4}$ sec using a Textronix oscilloscope, and then transferred to the computer for further analysis.
\begin{figure}[h]
\center
\includegraphics [width=8.5cm] {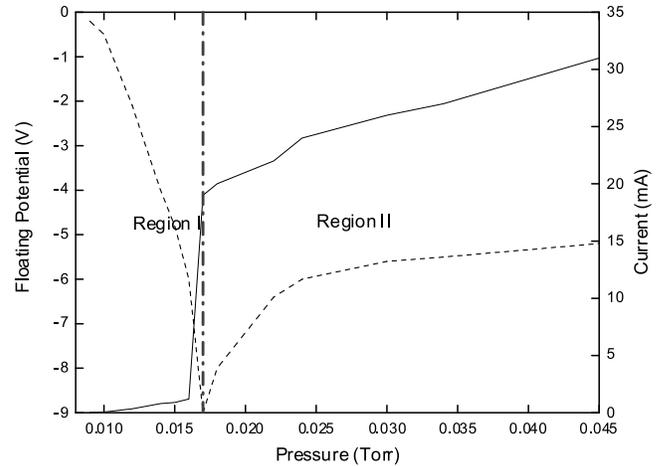}
\caption{The variation of plasma discharge current (solid line), and plasma floating potential (dotted line) with pressure. }
\label{fig:2}
\end{figure}
\begin{figure}[h]
\center
\includegraphics [width=8.5cm] {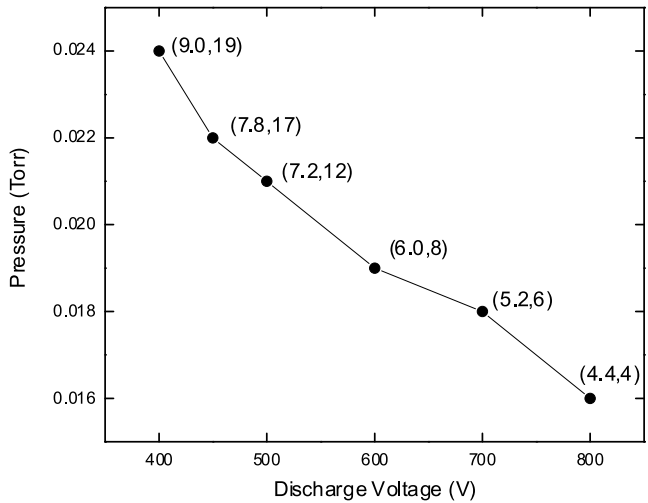}
\caption{Plot of critical pressures with discharge voltages. Corresponding floating potential, and discharge are within brackets. The first value in the brackets is the floating potential, and the second one is the discharge current in (V, mA) unit.}
\label{fig:3}
\end{figure}
\begin{figure}[h]
\center
\includegraphics [width=8.5cm] {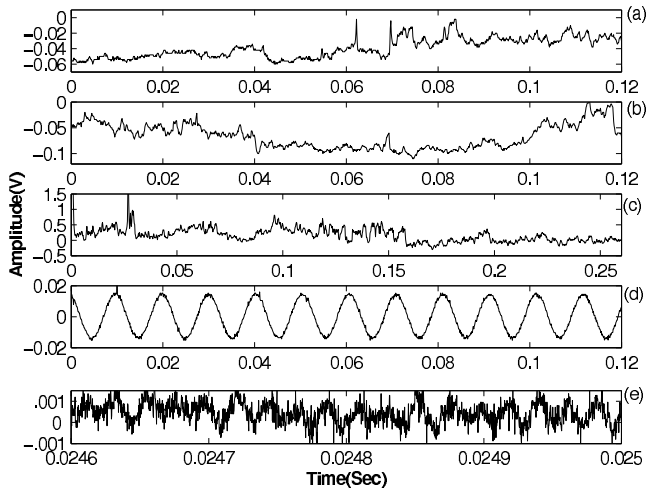}
\caption{Figure shows the electrostatic fluctuations at pressures $0.9\times10^{-2}$ (a), $1.2\times10^{-2}$ (b),  $1.5\times10^{-2}$ (c), $1.7 \times10^{-2}$  (d), and $2.2\times10^{-2}$ Torr (e) respectively.}
\label{fig:4}
\end{figure}

 Keeping the discharge voltage at a constant value of 800 V, a discharge was struck with a very faint glow at $\sim9\times10^{-3}$ Torr, and, then as the filling pressure was gradually increased by means of the needle valve, the intensity of the glow suddenly became bright at about $1.6\times10^{-2}$ Torr. The sudden change in the plasma dynamics at the critical pressure led to a change in the floating potential, and the discharge current.  The variation of  the plasma floating potential, and the plasma discharge current with pressure have been shown in Figure~\ref{fig:2}. It shows that, the current (solid line) jumped to a larger value by a factor of 15 at $1.6 \times 10^{-2}$ Torr, and then increased gradually with pressure.  On the other hand, The floating potential rapidly fell to a negative value ($\approx -9$ V) up to the same critical pressure ($1.6 \times 10^{-2}$ Torr), and then again increased (dotted line in Fig.~\ref{fig:2}) with further increase in pressure, until it finally settled down to $\approx -5$  V. So the critical pressure divides the scanned region into two regions I and II, shown by a vertical line ($-.-$)  in Figure~\ref{fig:2}.  The plasma density, temperature, and electron-electron collision mean free path in region II are $10^7-10^8cm^{-3}$, 2-4 eV, and $2.6\times10^5$ cm respectively. However, in  region I, it was almost impossible to obtain the I-V characteristics because of their extremely low values. Qualitatively looking at discharge current, and glow intensity, region II is probably a normal glow discharge region, while region I might be the dark or subnormal glow discharge region.  The critical pressure where the transition takes place is not a fixed point, but decreases with the discharge voltage as seen in  Figure~\ref{fig:3}. It is also seen that the floating potential, and the discharge current also decrease with discharge voltage. Figures~\ref{fig:4}(a), (b), and (c) are the typical electrostatic fluctuations at pressures $0.9\times10^{-2}$, $1.2\times10^{-2}$, and $1.5\times10^{-2}$ Torr respectively in region I, while (d), and (e) are the fluctuations at pressures $1.7 \times10^{-2}$ and $2.2\times10^{-2}$ Torr respectively in region II.
\section{Analysis of SOC behavior}
\label{sec:SOC}
The experimental evidences considered as main ingredients  of SOC are $1/f^{\beta}$ ($\beta>$0) power law (where $f$ is the frequency of the fluctuations obtained from Fast Fourier Transform) ~\cite{PRL:PBak,PhLettA:TLRhodes,PhyslettA:Skokov,PRL:Kim}, long-range correlation~\cite{POP:BACarreras2}, and nongaussian probability distribution function (PDF)~\cite{conf:Xu}. From the power spectral analysis we have estimated the $\beta$ from ln(Power) versus ln($f$). For long-range time correlation we estimated the Hurst exponent H, and the exponent ($\alpha$) of Autocorrelation function (ACF) decay, as described bellow.

\emph{\textbf{Hurst Exponent}}-The Rescaled-Range statistics $(R/S)$ method was proposed by Hurst and well established by Mandelbrot, and Wallis~\cite{POP:carreras1}. For the time series defined above, the $R/S$ is defined as~\cite{POP:carreras1,PhyslettA:Nirab} the ratio of the maximal range of the integrated signal normalized to the standard deviation:
\begin{equation}
\frac{R(n)}{S(n)}=\frac{max(0,W_1,W_2,...,W_n)-min(0,W_1,W_2,...,W_n)}{\sqrt{S^2(n)}}
\end{equation}
Here $W_k= x_1+x_2+x_3+...+x_k-k\overline{X(n)}$, where $\overline{X}$, $S^{2}(n)$, and n are respectively the mean, variance, and time lag of the signal. The expected value of $R/S$ scales like $cn^H$ as $n\rightarrow \infty$, where H is called the Hurst exponent. For random data H=0.5, while $H>0.5$  for the data with long range correlations. $H<0.5$ indicates the presence of long-range anti-persistency in the data. 

The ACF has been derived as follows:

\emph{\textbf{Auto-correlation}}-For a time series of length n, $X=[X_i,~i=1,2,...n]$, the ACF function can be written as ~\cite{PhysicaA:Davide}
\begin{equation}
C(\tau)=\frac{\frac{1}{n-\tau}\sum^{n-\tau}_{j=1}(X_{j+\tau}-\overline{X})(X_j-\overline{X})}{\frac{1}{n}\sum^n_{j=i}(X_j-\overline{X})^2}
\end{equation}
where $\overline{X}$, and $\tau$ are the mean, and time lag of the time series respectively.
If there is long-range time dependence in the system, then the algebraic decay of the ACF can be written as ~\cite{PRE:GRangarajan}
  
\begin{equation}
C(\tau)\sim \tau^{-\alpha}
\end{equation} 
for large $ \tau$, where $0<\alpha<1$. 

In order to verify nongaussianity we obtained the PDF of the fluctuating data.

\begin{figure}[h]
\center
\includegraphics [width=8.5cm]{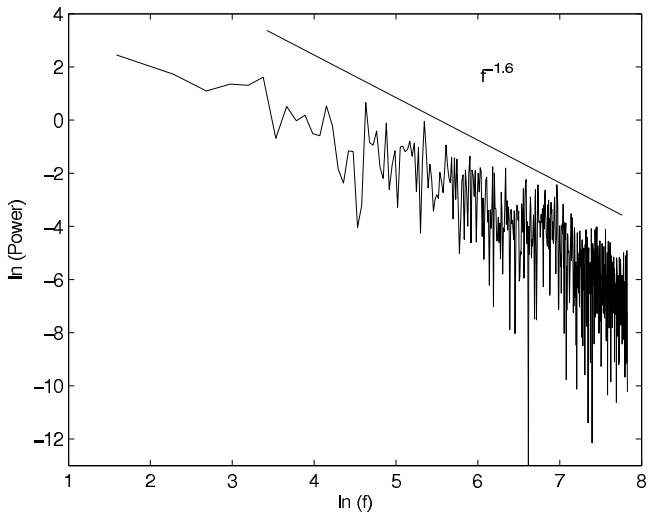}
\caption{ln(power) vs. ln $f$ plot. The solid line shows the best fit.}
\label{fig:5}
\end{figure}
\begin{figure}[h]
\center
\includegraphics [width=8.5cm] {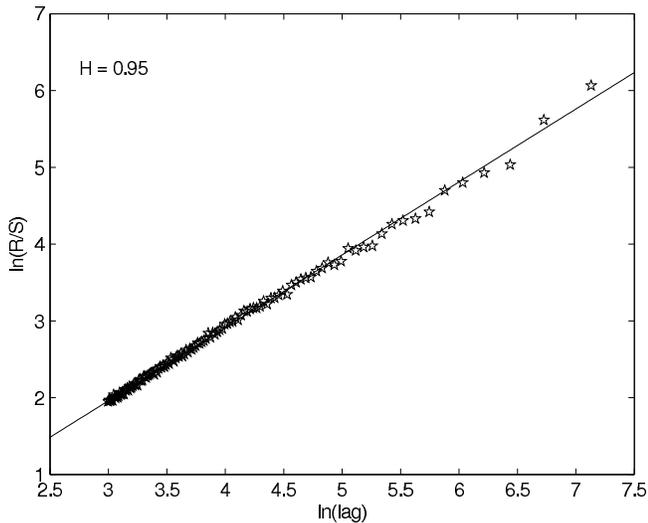}
\caption{ln(R/S) vs. ln(lag) plot for the electrostatic fluctuation at the pressure $1.4\times10^{-2}$ Torr. The solid line shows best fit.}
 \label{fig:6}
\end{figure}
\begin{figure}[h]
\center
\includegraphics [width=8.5cm] {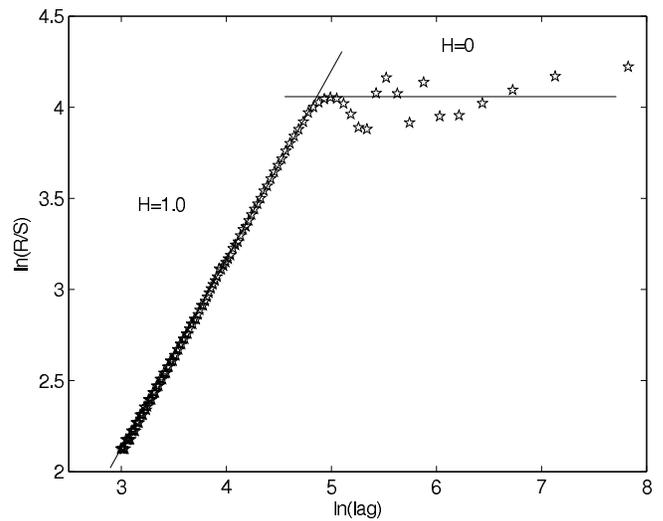}
\caption{R/S as a function of time lag  for the electrostatic fluctuation of the coherent oscillations at the pressure $1.7\times10^{-2}$ Torr. For one time period lag, H=1, and for lag more than one period, H=0.}
\label{fig:7}
\end{figure}
\begin{figure}[h]
\center
\includegraphics [width=8.5cm] {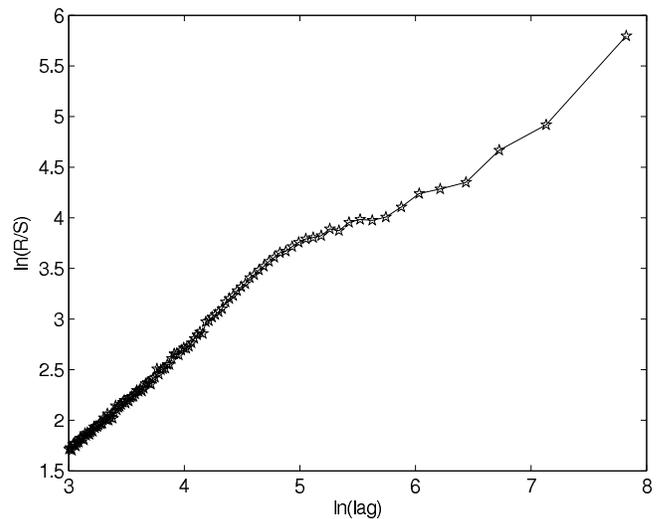}
\caption{ln(R/S) versus ln(lag) plot at the pressure $2.2\times10^{-2}$ Torr, more than one slope indicates instabilities with many modes.}
\label{fig:8}
\end{figure}

\begin{figure}[h]
\center
\includegraphics [width=8.5cm] {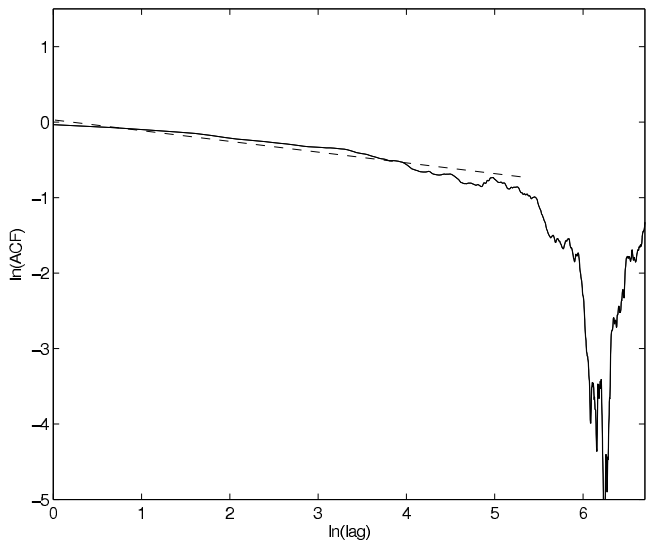}
\caption{ln-ln plot of ACF vs. time lag at pressure $1.4\times10^{-2}$ Torr. Up to 6 decorrelation times, it shows power law (dotted line), and after that it follows exponential decay.}
\label{fig:9}
\end{figure}
\begin{figure}[h]
\center
\includegraphics [width=8.5cm] {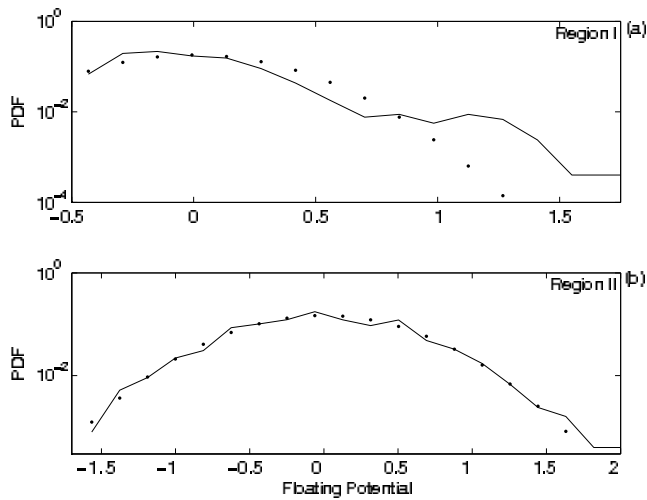}
\caption{The probability distribution function (PDF) of the fluctuation at pressure $1.2\times10^{-2}$ Torr, and the dotted line is the corresponding gaussian fit. }
\label{fig:10}
\end{figure}

Figure~\ref{fig:5} shows ln (Power) vs. ln $f$ of the fluctuations in region I from which we estimated the exponent to be $\approx 1.6$.  This  agrees well with the numerical ~\cite{PRL:PBak,PRA:PBak}, as well as experimental observations ~\cite{PhLettA:TLRhodes,PhyslettA:Skokov,POP:BACarreras2} in the presence of SOC behavior.

Figure~\ref{fig:6} shows a typical plot of ln(R/S) vs. ln(time lag) of the fluctuations in region I, for a pressure of $1.4\times10^{-2}$. The Hurst exponent H is about $0.96\pm.01$ (this indicates long-range time correlations)~\cite{PRL:BACarreras00}, and it is almost constant over the entire pressure range in region I.  On the other hand, for sinusoidal like oscillation in region II, the Hurst exponent is 1 for the lag length of one period of oscillations, and zero with more than one period lag~\cite{POP:carreras1} as shown in Figure~\ref{fig:7}. Also in the same region (II) multi slope ln(R/S) vs. ln(lag) plot as seen in Figure~\ref{fig:8} probably due to plasma instabilities of many frequencies~\cite{POP:carreras1}. The ACF exponent $\alpha$ has been calculated for the fluctuations in region I from the ln(ACF) vs. ln(time lag) plot as shown in Figure~\ref{fig:9}. The ACF (Fig.~\ref{fig:9}) shows power law up to about 6 times the decorrelation time, and after that it follows exponential decay. Average value of $\alpha$ is about 0.30. Using the relation $H=(2-\alpha)/2$ ~\cite{PRE:GRangarajan}, H calculated from ACF is $\sim$ 0.85, which is close to the value of H, calculated using R/S technique. In region II no power law decay of ACF has been observed.
\begin{table}[h!b!p!]
\center
\caption{In the following table the decay exponent $\alpha$ of the ACF, H from ACF, H using R/S, and the power spectral index $\beta$ have been shown in the second, third, fourth, and fifth column respectively, for the pressures shown in the column one. }
\begin{tabular}{c c c c c}
\hline
pressure & $\alpha$  & Hurst   & Hurst  & $\beta$  \\
$ 10^{-2}$(Torr) & (ACF)& (ACF)  & (R/S) & (PS)\\
\hline
\hline
0.9 & 0.38 & 0.81 & 0.97 & 1.85\\
1.2 & 0.34 & 0.83 & 0.96 & 1.65\\
1.5 & 0.23 & 0.88 & 0.95 & 1.60\\
\hline
\end{tabular}
\label{table1}
\end{table}
The PDF of the floating potential fluctuations in region I seen in Fig.~\ref{fig:10}(a) clearly shows a nongaussian nature. Corresponding best gaussian fit is given by dotted curve in the same figure. We suspect that there might be a slightly bimodal distribution similar to Ref.~\cite{PhyslettA:Skokov}.  Figure~\ref{fig:10}(b) shows the Gaussian nature of the fluctuations in region II.

Our results of Hurst exponent, $H>0.5$, ACF exponent, $\alpha\sim0.30$, nongaussian PDF, and power spectral index $\beta \sim 1.60$ in the pressure range $9\times10^{-3}-1.6\times10^{-2}$ Torr, are consistent with the systems exhibiting SOC like behavior.

Comparison of $\alpha$, H by ACF, H by R/S, and $\beta$ have been shown in  Table ~\ref{table1}, for pressures $0.9\times10^{-2}$, $1.2\times10^{-2}$, $1.5\times10^{-2}$ Torr.
\section{Conclusion}
\label{sec:conclusion}
We have obtained SOC behavior over a finite range of neutral pressure of $ 9\times10^{-3}-1.6\times10^{-2}$ Torr for a fixed discharge voltage. Glow discharges are simple systems, but their physics can be quite complicated due to the presence of several phenomena like avalanche breakdown, ionization waves, low frequency ion-acoustic instability, double layer, chaos etc. Most of them are highly nonlinear processes and hence one requires different techniques both statistical and spectral to investigate and understand their behavior. From our present analysis we observe that the plasma dynamics in the region I is compatible with self organized criticality, while region II is not. This could also imply that plasma transport in region I is quite different from region II. Detailed investigations of the spatio-temporal chaos, and the multifractal nature of these fluctuations are in progress and will be reported else where.
\section*{Acknowledgment}
We would like to thank the Director SINP, for his constant support and A. Bal, A. Ram, S. Sil, D. Das, and M. Chattopadhya of Plasma Physics Division for the technical assistance, and the other members for their encouragement whenever required.

\end{document}